\documentclass[aps,prl,twocolumn,groupedaddress]{revtex4}
\usepackage {amssymb}
\usepackage {amsmath}
\usepackage {graphicx}
\usepackage {longtable}

\begin{document}
\title{Conductivity behavior of La$_{0.75}$Ca$_{0.25}$MnO$_{3}$ in vicinity of
ferromagnetic-paramagnetic transition studied with single current
pulses}
\author{V. B. Krasovitsky}
\affiliation{B. Verkin Institute for Low Temperature Physics and
Engineering, National Academy of Sciences, prospekt Lenina 47,
Kharkov 61103, Ukraine}

\author{B. I. Belevtsev}
\email[]{belevtsev@ilt.kharkov.ua}
\affiliation{B. Verkin Institute for Low Temperature Physics and
Engineering, National Academy of Sciences, prospekt Lenina 47,
Kharkov 61103, Ukraine}

\begin{abstract}
Temperature and current dependences of resistivity of bulk
La$_{0.75}$Ca$_{0.25}$MnO$_{3}$ sample grown by the floating-zone
method were studied using single ramp pulses of current. It is
found that near the Curie temperature $T_C$ the sample resistance
depends substantially on current magnitude. The observed features
can be determined by inhomogeneous Joule overheating due to mixed
phase state of manganites near the ferromagnetic-paramagnetic
transition and percolation character of this transition.

\end{abstract}
\maketitle

 In recent years extensive studies were done on complex
manganese oxides with the perovskite structure which are known as
manganites with the general formula R$_{1-x}$A$_{x}$MnO$_{3,}
$where R is a rare-earth element, A an alkaline-earth element. The
keen interest in the manganites is accounted for by their unique
physical properties and possibilities of their practical
applications. One of those properties is their colossal (negative)
magnetoresistance near the Curie temperature $T_{C}$ of the
transition from paramagnetic (PM) to ferromagnetic (FM) states.
This transition is often accompanied by such strong (sometimes,
tenfold) reduction of the electrical resistance that it looks like
an insulator-metal transition. The magnitude of this effect
depends on the crystal perfection of samples, their chemical
composition, while being very sensitive as well to external
influence, in particular to magnetic and electric fields. In the
well-known studies on influence of the electric field (or the
current J) in bulk and film samples of the manganites (for
example, \cite{sud, gao,palan,imam}) the observation was made, that, to
some extent or another, the Joule selfheating of samples by the
measuring current is always present. This phenomenon may well result in thermal
bistability in such metals where in the phase transition region a
stepwise resistance-on-temperature dependence is observed with the
positive temperature coefficient \cite{gurev}.
\par
The present study concerns with the non-linear effect in
conductivity of a bulk specimen of La$_{0.75}$Ca$_{0.25}$MnO$_{3}
$ grown by the floating-zone technique with the radiation heating
\cite{boris1,boris2}. The sample structure was close to being
single-crystal. For the purpose of measurements such part of the
sample was used which had been studied earlier in reference
\cite{boris1}. It is known \cite{boris1,boris2,boris3} that the
PM-FM transition in this compound is first order and has a
percolation character. In the region of this transition the system
is inhomogeneous and consists of a mixture of PM and FM phases.
The non-Ohmic effects in these systems are attributed, first of
all, to the Joule heating which raises the temperature of the
specimen. However, in the inhomogeneous systems the Joule heating
may be inhomogeneous as well. This study aims to reveal such
effects. For this purpose, the behavior of the specimen
conductivity in the PM-FM transition region was studied by
applying single ramp pulses of current that are
linearly-increasing with  time (in the quasi-adiabatic regime).
Thus we expected to be able to keep a global heating of the
specimen not too strong (as compared to the well-known DC
studies), while simultaneously providing for a large enough value
of the measuring current which can give a possibility of
determining the peculiarities of the electron transport in this
inhomogeneous medium.
\par
The measurements were taken in the temperature range 77-320 K. The
specimen ($2\times 2\times 3$ mm$^{3}$) was glued to a copper
plate (coated with a thin insulating intermediate layer) which was
placed in a chamber of highly rarefied gaseous helium (10$^{-3}$ -
10$^{-4}$ Torr), immersed in liquid nitrogen. The current and
potential leads to the specimen were soldered with indium. The
electric measurements were carried out using a standard four-probe
technique applying single current pulses that were linearly
increasing with time (``sawtooth'' shape) with the widths t$_{p}$
ranging from 0.25 to 5 ms and an amplitude J$_{max}$ up to 10 A
(the maximum current density being about 2.5$\times $10$^{2}$
A/cm$^{2}$). The operating range of the current pulse widths was
chosen such that the width t$_{p}$, on the one hand should exceed
the characteristic times of the magnetic relaxation of the
manganite samples of this kind and, on the other hand, that it
should meet the desirable minimization of the heat release on this
specimen. The latter is determined not so much by the current
pulse width, as by its amplitude and the resistivity value of the
specimen which during the PM-FM transition (in the temperature
interval 200-230 K) varies $\sim $40-fold (Fig.1). Upon
cooling-down of the specimen to the liquid nitrogen temperature
the measurements were taken at fixed temperatures, the voltage
drop response on the specimen $U(t)$ and signal, that was
proportional to the magnitude of current passing through the
specimen $J(t)$, being recorded simultaneously at a dual-beam oscilloscope
``V6-9''. Fig. 1 (inset) presents two types
of the characteristic oscillograms $U(t)$ and $J(t)$. In one case,
the curves of current and voltage are self-similar (type 1, Fig.1,
inset ``a''). In the other case (in the steep resistance increase
region during the PM-FM transition) the curves $U(t)$ display two
clear-cut turns (type 2, Fig.1, inset ``b'').
\par
The  temperature dependences of resistance at assigned current
$R_{J}(T)$ and the current dependences  of resistance at assigned temperature
$R_{T}(J)$ were determined from data of the corresponding
oscillograms $U(t)$ and $J(t)$. The construction of the relations
$R_{J}(T)$ involved the use of current pulses of varying widths
and amplitude. To obtain the base relation $R_{0}(T)$ that would
be undistorted by the current influence (Fig.1) such data was processed that
pertained to short current pulses of low amplitude
$t_{p}=0.25$~ms; $J_{max}=1$~A. At those parameters, the
oscillograms of the first type were realized (Fig.1) which was
indicative of the fulfillment of the Ohm law. The obtained
relation $R_{0}(T)$ coincides with the temperature dependence
$R(T)$, measured on the same specimen \cite{boris1} at low DC
current below 5 mA.
\par
Fig.2 shows the relations $R_{J}(T)$ for three current values of
2.7, 5.4 and 8 A, which were realized accordingly in 0.25, 0.5 and
0.75 ms after triggering the pulse. At any temperature the Joule
heating causes the specimen to raise its temperature. The relation
$R(T$) in the Ohmic regime is a curve with a rather steep maximum
at $T_{p}\approx 230$~Ê (Fig.~1). It is clear that at $T < T_{p}$
the Joule heating should increase the resistance, and at values
higher than $T_{p}$ it should bring it down, which is what is
observable in the first approximation. The inset of Fig.~2
presents the temperature dependence of a relative resistance
change during the variations of the current $\Delta R_{J} \left(
{T} \right)/R_{J_{0}}  \left( {T} \right) = \left[ {R_{J} \left(
{T} \right) - R_{J_{0}}  \left( {T} \right)} \right]/R_{J_{0}}
\left( {T} \right)$, where $R_{J_{0}}  \left( {T} \right)$ -
resistance measured at the minimum current. The characteristic
feature of it is such that the peak $\Delta R_{J} \left( {T}
\right)$ is in the region of the sharpest resistance growth during
the magnetic transition (about 210 K).
\par
Major peculiarities of the non-linear behavior came up with the
resistance-current relations $R_{T}(J)$ being in the temperature
range 200--240 K (Fig.~3). Below the temperatures $ \approx
203-205$ K the resistance of the specimen does not depend on the
current (in the current range studied). Above the temperatures
$ \approx 214-215$ K the resistance increases with rising current
(type 1 oscillogram, Fig.1, inset ``a''), while above $ \approx
220$~K it increases at low currents just to come down for
relatively large currents, which is consistent with the data
presented in Fig.~2. All of this is in line with the anticipated
Joule heating influence for approximately homogeneous systems (see
similar results for the system La-Ca-Mn-O \cite{sud,gao}).
\par
Yet, in the interval $ \approx 205-214$~K, the curves of
$R_{T}(J)$ demonstrate an absolutely different nature of the
non-linearity featuring two turns which are characteristic of the
type 2 oscillogram (Fig.1, inset ``b''). This behavior is
inconsistent with the Joule heating influence for homogeneous
systems and it must be attributed to certain peculiarities of the
magnetic transition of first order under study. In this
temperature interval of the magnetic transition, the system is
inhomogeneous to the utmost, consisting of a mixture of PM and FM
phases \cite{boris1,boris2,boris3}. The first turn which is
observable at the smallest values of the current (empty circles in
Fig.3) is obviously associated with an incipient noticeable
influence of the Joule heating of the specimen.  The second turn
on the curve $R_{T}(J)$ (Fig.3, filled circles) is observable for
far greater currents and, as much as the first turn, it is shifted
toward lesser currents with increasing temperature. The average
specimen temperature increase as observable by emergence of the
second turn can be determined by following the resistance change
values from initiation of the pulse to this turn, while comparing
them with the relation $R_{J}(T)$ at the minimum current. In the
temperature interval 210--214 K this variation is about
0.1~$\Omega$, which corresponds to the specimen being heated by
only 2 K. The appropriate estimation of the specimen heating made
in the assumption of the adiabatic nature of the process according
to the formula $\Delta T = Q_{R} /mC$ ($m$ and $C$ - mass and heat
capacity of the specimen) comes out with a closely matched value.
Note that the resistance of the current contacts at all
temperatures is substantially smaller than that of the specimen
and cannot be the cause for its heating. This statement is
supported by the specimen resistance being independent of the
value of the current at temperatures below $ \approx 203$~K, which
is to say when the specimen resistance is smaller than several
hundredths of one Ohm.
\par
The specimen studied in this research had been investigated in
reference \cite{boris1} where DC resistance, magnetization,
velocity and absorption of the sound were measured. The anomalies
of these properties observed in reference \cite{boris1} in the
vicinity of the PM-FM transition were interpreted in terms of the
existence of a multi-phase state near $T_{C}$. It is well-known
\cite{boris2,boris3} that in this system the FM clusters may be
present at temperatures far above $T_{C}$, while in the meantime a
certain number of PM dielectric clusters can be preserved at
temperatures that are considerably lower than $T_{C}$. This
situation implies truthfulness of the percolation nature of the
FM-PM transition. This has served as basis, as well, for the
interpretation of results of this work. During temperature
increase near $T_{C}$ the volume portion of the FM phase
decreases, while the PM one is on the increase. Near the
percolation threshold, the conductivity of a system is determined
by a small number of high-conductive FM percolation channels
(paths). Under those conditions an inhomogeneous Joule heating is
possible which is made feasible by the supply of short current
pulses so that the locally generated heat has not time enough to
dissipate in the ambience during the time of the measurements.
Earlier, an inhomogeneous Joule heating was observable at DC
measurements in thin and narrow films of
La$_{0.7}$Ca$_{0.3}$MnO$_{3}$ nearly 50 $\mu$m wide \cite{palan},
which allowed for considerable current densities to be achieved.
In this work, we observed jumps of resistance after exceeding the
values of the current of some critical magnitude (which were
temperature-dependent). Those kinds of jumps are characteristic of
the appearance of a bistable state in the region of the phase
transitions with S-shaped $V$-$I$ characteristics \cite{gurev}
(which was observable as well in manganite films \cite{imam}). The
unstable states of such kinds that are observable at large
currents are quite feasible for the homogeneous systems, as well.
The inhomogeneity which is inherent to the manganites in the
region of their phase transition enhances such kinds of effects.
Really, a large local overheating in the region of the percolation
FM current paths causes, at least, a part of these paths to
transition to the PM state so that instead of a continuous channel
there appears a chain of isolated FM islands in the dielectric PM
matrix. This brings about a dramatic increase in the resistance of
the entire system which is reflected in the type 2 oscillograms
(Fig. 1, inset ``b''). In this way, the
pulse-application technique has enabled to discover the appearance
of an inhomogeneous percolation structure in a bulk specimen of
the manganite near the phase transition which agrees well with
results from other studies \cite{boris1,boris2,boris3}.

\newpage

\begin{figure}[tb]
\centering\includegraphics[width=0.95\linewidth]{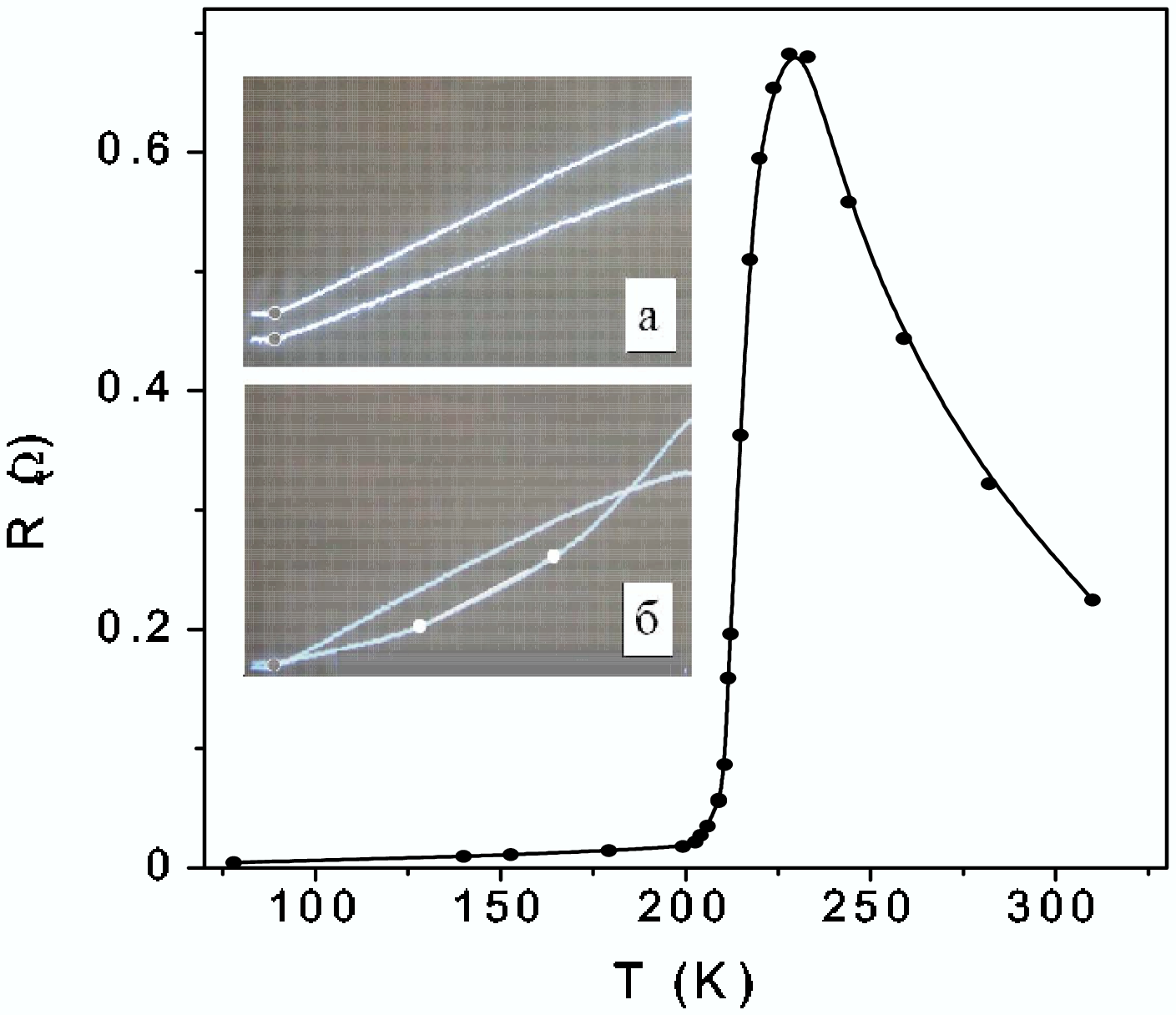}
\caption{Temperature variations of specimen resistance in the
Ohmic regime. In the inset: an example shown in the oscillograms
of the drop of voltage (uppermost curves) and current (lowermost
curves), t$_{p}$=3.6 ms and J$_{max}$ = 8 A, type 1 (a) and type 2
(b).}
\end{figure}

\begin{figure}[htb]
\centering\includegraphics[width=0.95\linewidth]{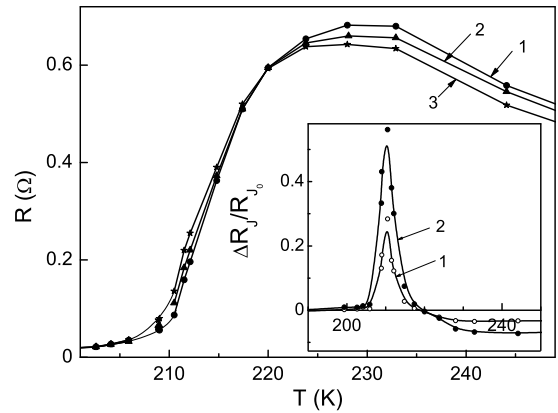}
\caption{Dependence of resistance of specimen on temperature for
currents 2.7, 5.4 and 8 A (Curves 1, 2 and 3, respectively). In
the inset: the temperature dependences $\Delta R_{J} \left( {T}
\right)/R_{J_{0}}  \left( {T} \right) = \left[ {R_{J} \left( {T}
\right) - R_{J_{0}}  \left( {T} \right)} \right]/R_{J_{0}} \left(
{T} \right)$ (see text). 1 - $J=5.4$~A, 2 - $J=8$~A.}
\end{figure}

\newpage
\begin{figure}[ht]
\centering\includegraphics[width=1.00\linewidth]{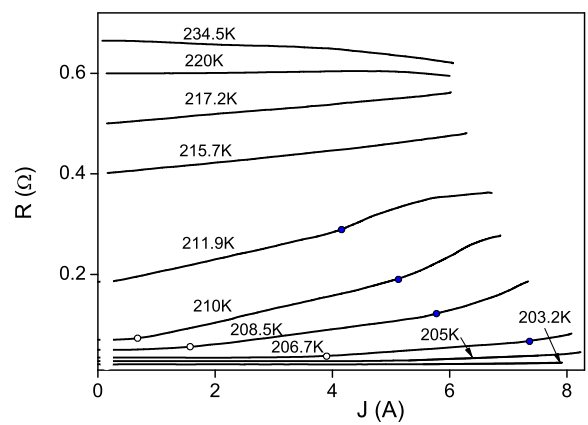}
\caption{Dependence of resistance of specimen on current for
different temperatures.}
\end{figure}

\end{document}